# Uncertainty-Aware Structure–Property Mapping of Spinodoid Metamaterials *via* Heteroscedastic Gaussian Process Regression

Minwoo Park[a]†, Junseo Park[a]†, Mingyu Lee[a]†, Hugon Lee[a], Hanbin Cho[a], Ikjin Lee[abc]*, and Seunghwa Ryu[abc]*

[a]Department of Mechanical Engineering, Korea Advanced Institute of Science and Technology (KAIST), Daejeon 34141, Republic of Korea

[b]KAIST InnoCORE PRISM-AI Center, Korea Advanced Institute of Science and Technology (KAIST), Daejeon 34141, Republic of Korea

[c]Department of AX, Korea Advanced Institute of Science and Technology (KAIST), Daejeon 34141, Republic of Korea

†These authors have contributed equally: Minwoo Park, Junseo Park, and Mingyu Lee

*Corresponding author. E-mail: ryush@kaist.ac.kr, and ikjin.lee@kaist.ac.kr

## Abstract

Spinodoid metamaterials offer a broad, tunable design space for anisotropic mechanical properties, yet their structure–property relationships are commonly treated as representative mappings from cone-angle descriptors to single effective stiffness values. This deterministic view overlooks the stochastic nature of Gaussian random field (GRF)-based topology generation, where identical cone-angle descriptors can produce different morphology realizations and property scatter. Here, we present an uncertainty-aware structure–property mapping framework that reinterprets cone-angle descriptors as stochastic descriptors associated with input-dependent property distributions. Using heteroscedastic Gaussian process regression (GPR), the framework infers input-dependent predictive uncertainty from sparse one-realization-per-point data without requiring empirical variance labels at every design point. The results show that stiffness scatter differs across tensor components according to each component's mechanically active directions, and that parameter sets yielding identical mean stiffness can carry different aleatoric uncertainty. Applying this uncertainty to reliability-based design optimization (RBDO), we show that a deterministic optimum is highly susceptible to constraint violation once morphology-induced variability is considered, and that a homoscedastic RBDO formulation fails to meet the prescribed reliability target — only the heteroscedastic formulation satisfies the reliability target under the heteroscedastic uncertainty evaluation. This establishes uncertainty-aware surrogate modeling as essential for reliability-aware inverse design of spinodoid metamaterials; extending the framework to nonlinear responses remains for future work.

# 1. Introduction

Mechanical metamaterials have been widely investigated as material systems that can realize stiffness[1–7], anisotropy[8–11], and energy absorption[12,13] that are difficult to achieve in conventional materials through the geometric design of their microarchitectures. In particular, spinodoid metamaterials[14,15] have non-periodic topologies and smooth morphologies inspired by spinodal decomposition, offering the advantage of mitigating sharp junctions and stress concentrations that often arise in truss- or plate-based lattices[16,17]. In addition, because their topologies can be efficiently generated based on Gaussian random fields (GRFs)[18], a broad design space and tunable anisotropy can be explored through relative density and cone-angle parameters without directly performing costly phase-field simulations[19,20]. Based on these characteristics, previous studies[14,15,21–24] have combined computational homogenization and machine-learning surrogates to establish structure–property mappings between cone-angle parameters and effective stiffness tensors, and have extended these mappings to inverse design of target anisotropic stiffness and functionally graded spinodoid structures.

In existing spinodoid design frameworks[14,15,21–25], a given design parameter set has commonly been associated with a representative effective property for practical efficiency in surrogate modeling and inverse design. In other words, when relative density and cone-angle parameters are specified, a representative effective stiffness is defined for the corresponding parameter set, and a representative structure–property mapping is learned based on this relation. Such a deterministic representative mapping is a useful simplification for efficiently exploring the broad spinodoid design space and performing target-property-based inverse design. However, from the perspective of GRF-based stochastic generation, a cone-angle parameter set does not directly specify a single morphology, but rather defines an ensemble of possible morphology realizations. As illustrated in Fig. 1(a), even under the same cone-angle parameter set, different

GRF realizations can lead to different local morphologies, connectivities, and surface arrangements.

This morphology variability can, in turn, lead to variability in effective stiffness. That is, for a given cone-angle descriptor $\Theta$, effective stiffness can be interpreted not as a single deterministic property value, but as a realization-dependent quantity governed by the GRF realization. From this perspective, the structure–property relationship can be extended from a representative mapping, $\Theta \to \mathbb{C}^{\text{eff}}$, to a distributional mapping, $\Theta \to p\left(\mathbb{C}^{\text{eff}} \middle| \Theta\right)$. Fig. 1(b) illustrates this concept. Different morphology realizations can produce different effective stiffness responses, and consequently, a property distribution can arise even under the same design parameter set.

Furthermore, the width of this property distribution may not be uniform across the design space. In some cone-angle regions, changes in GRF realization may strongly affect load-bearing pathways and lead to large property scatter, whereas in other regions, effective stiffness may remain relatively stable even when the morphology realization changes. Fig. 1(c) conceptually illustrates that this realization-induced property variability can depend on the cone-angle input. Therefore, the spinodoid structure–property relationship may be more appropriately interpreted in terms of input-dependent noise, or heteroscedastic behavior, rather than constant noise. In this study, we first refer to this physical variability as realization-induced property variability and then interpret it, from the perspective of surrogate modeling, as morphology-induced aleatoric uncertainty that is not fully explained by the cone-angle descriptor alone.

However, directly evaluating such property distributions over the full design space is computationally impractical. Obtaining the true property distribution at each cone-angle parameter set would require sufficiently many GRF realizations and corresponding computational homogenization analyses[14,26–28], which is prohibitive when repeated across possible design points.

Therefore, this study does not use empirical variance as a directly supervised target at every design point. Instead, repeated GRF realizations are generated at selected design points to diagnostically verify the presence of realization-induced property scatter, while a surrogate modeling strategy is used over the broader design space to infer input-dependent predictive uncertainty from sparse one-realization-per-point observations.

This study proposes an uncertainty-aware structure–property mapping framework for spinodoid metamaterials generated by stochastic GRFs. First, we examine whether identical cone-angle descriptors can generate multiple morphology realizations and corresponding property scatter. Second, we employ heteroscedastic Gaussian process regression (GPR)[29] to infer input-dependent predictive uncertainty from limited one-realization-per-point simulation data, while distinguishing it from epistemic uncertainty due to sparse sampling. Homoscedastic GPR[30,31] is used only as a baseline in the controlled one-dimensional analysis. Third, the calibrated surrogate is applied to deterministic design optimization (DDO) and reliability-based design optimization (RBDO), demonstrating that a deterministic optimum is highly susceptible to constraint violation when morphology-induced variability is accounted for, and that a homoscedastic uncertainty assumption can fail to meet prescribed reliability targets when stochasticity varies across the design space. Through this framework, spinodoid design parameters are reinterpreted as stochastic descriptors associated with property distributions, providing a physically consistent basis for reliability-aware inverse design of spinodoid metamaterials.

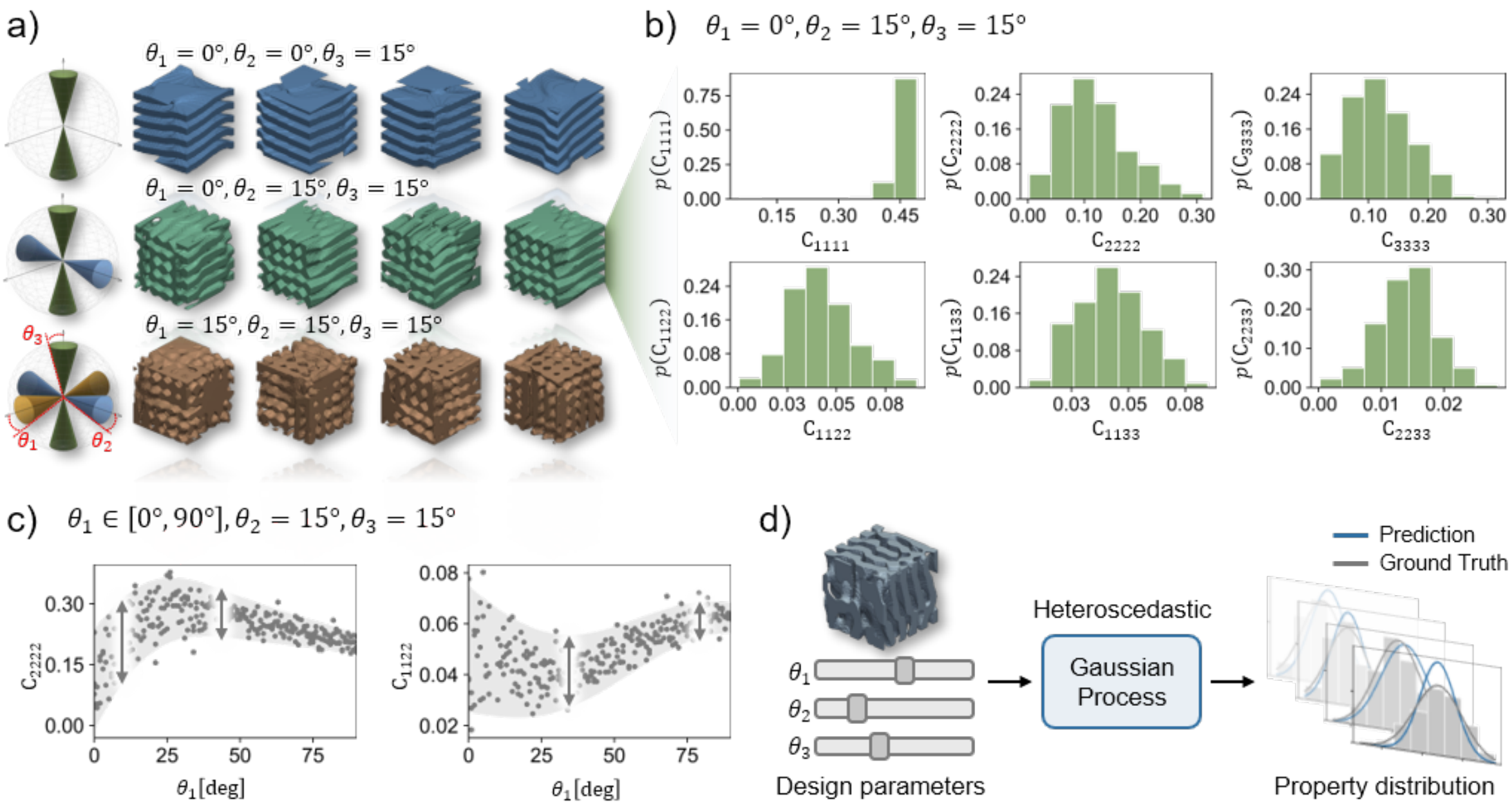

Fig. 1. Conceptual framework for uncertainty-aware spinodoid structure–property mapping. (a) Multiple morphology realizations generated from the same cone-angle parameter set. (b) Corresponding property distributions for selected stiffness components. (c) Input-dependent scatter along a controlled cone-angle path. (d) Heteroscedastic GPR framework for inferring input-dependent predictive uncertainty from sparse observations.

## 2. Generation of spinodoid topologies

Spinodal topologies are smooth, bicontinuous, and non-periodic morphologies inspired by the concentration patterns formed during spinodal decomposition[19,20]. In this study, spinodoid topologies were generated using a Gaussian random field (GRF)-based stochastic construction, which approximates the early-stage concentration fluctuations of spinodal decomposition as a superposition of standing waves. This construction avoids directly solving the time-dependent Cahn–Hilliard phase-field equation while retaining a compact parametrization of anisotropic spinodal-like architectures[21,22,25]. Each spinodoid morphology is defined from a scalar phase field,

$$\varphi(\mathbf{x};\Theta,\omega)=\sqrt{\frac{2}{N}}\sum_{i=1}^{N}\cos(\beta\mathbf{n}_i\cdot\mathbf{x}+\gamma_i) \tag{1}$$

where $\mathbf{x}\in\Omega\subset\mathbb{R}^3$ is the spatial coordinate in the representative volume element (RVE), $N\gg 1$ is the number of standing waves, $\beta>0$ is the characteristic wavenumber controlling the microstructural length scale, $\mathbf{n}_i\sim\mathcal{U}(S^2)$ is the unit wave-vector direction, and $\gamma_i\sim\mathcal{U}([0,2\pi])$ is the phase angle. The design descriptor is defined as $\Theta=(\theta_1,\theta_2,\theta_3)$, where $\theta_1$, $\theta_2$, and $\theta_3$ are the cone-angle parameters. In this study, the relative density, RVE size, spatial resolution, and $\beta$ were fixed, so that the following analyses focus on the effect of the cone-angle descriptor.

Anisotropy is introduced by restricting the admissible wave-vector directions to a cone-angle-defined subset of the unit sphere, $S_\Theta\subset S^2$:

$$S^2=\{\mathbf{k}\in\mathbb{R}^3:||\mathbf{k}||=1\} \tag{2}$$

$$S_\Theta=\{\mathbf{k}\in S^2:|\mathbf{k}\cdot\hat{\mathbf{e}}_1|>\cos\theta_1\vee|\mathbf{k}\cdot\hat{\mathbf{e}}_2|>\cos\theta_2\vee|\mathbf{k}\cdot\hat{\mathbf{e}}_3|>\cos\theta_3\} \tag{3}$$

where $\hat{\mathbf{e}}_1$, $\hat{\mathbf{e}}_2$, and $\hat{\mathbf{e}}_3$ are the Cartesian bases. The condition $|\mathbf{k}\cdot\hat{\mathbf{e}}_j|>\cos\theta_j$ defines a double-cone region around the $\pm\hat{\mathbf{e}}_j$ directions. A smaller cone angle imposes stronger

directional constraints on the GRF wave vectors, whereas a larger cone angle permits a broader orientation distribution. Because the generated interfaces tend to align preferentially perpendicular to the sampled wave vectors, the cone-angle descriptor $\Theta$ directly controls the anisotropy of the resulting topology.

For a prescribed $\Theta$, a single GRF realization $\omega = \{(\mathbf{n}_i, \gamma_i)\}_{i=1}^{N}$ specifies the particular wave-vector directions and phase angles used to construct one morphology instance. $\Theta$ therefore controls the *admissible* class of morphologies through the orientation distribution of the wave vectors, whereas $\omega$ determines the *specific morphology* realized within that class — so that resampling $\omega$ under the same $\Theta$ yields a different morphology, as illustrated in Fig. 1(a).

The generated phase field was converted into a solid–void topology by thresholding. For a prescribed relative density $\rho$, the threshold $\varphi_0$ is chosen such that the solid volume fraction matches $\rho$:

$$\chi(\mathbf{x}; \Theta, \omega) = \begin{cases} 1, & \varphi(\mathbf{x}; \Theta, \omega) \leq \varphi_0 \\ 0, & otherwise \end{cases} \tag{4}$$

where $\chi = 1$ and $\chi = 0$ denote the solid and void phases, respectively. The effective stiffness of each generated topology is then evaluated by computational homogenization. Each RVE is treated as composed of a homogeneous, isotropic, linear elastic base material with Young's modulus $E_s = 1.0$ and Poisson's ratio $\nu_s = 0.3$. Affine displacement boundary conditions were applied to compute the effective fourth-order elastic stiffness tensor:

$$\mathbb{C}^{\text{eff}}(\Theta, \omega) = \mathcal{H}[\chi(\mathbf{x}; \Theta, \omega)] \tag{5}$$

where $\mathcal{H}$ denotes the computational homogenization operator. This notation explicitly reflects that the effective stiffness depends on both the design descriptor $\Theta$ and the stochastic realization $\omega$ — not on $\Theta$ alone.

Two design-space scenarios are considered in the following analyses. The first is a controlled one-dimensional path, $\theta_1 \in [0°, 90°], \theta_2 = 15°, \theta_3 = 15°$, which isolates the effect of a single cone-angle parameter while holding the other directional constraints fixed. The second is the full three-dimensional space, $(\theta_1, \theta_2, \theta_3) \in [0°, 90°]^3$, used to evaluate the generalization of uncertainty-aware structure–property prediction across the complete design space.

## 3. Heteroscedastic GPR for uncertainty-aware structure–property mapping

Following the distributional structure–property interpretation introduced in Section 2, we model the relation between the cone-angle descriptor $\Theta$ and a selected effective stiffness component $y$ probabilistically. Here, $y$ denotes one component of the homogenized stiffness tensor, such as $C_{1111}$, $C_{3333}$, or $C_{2323}$. Because each spinodoid topology is generated from a stochastic GRF realization, an observed stiffness value can be written as $y = C_k(\Theta, \omega)$ (see Eq. 5). Although the conditional response $p(y|\Theta)$ is the quantity implied by the stochastic topology generation process, directly estimating this distribution at every design point would require repeated GRF generation and computational homogenization throughout the full cone-angle space. We therefore use heteroscedastic GPR to infer input-dependent predictive uncertainty from sparse one-realization-per-point observations — a training protocol in which each design point $\Theta_i$ is associated with exactly one GRF realization $\omega_i$ and its corresponding stiffness value, without computing empirical variance at any design point.

The observation model is written as

$$y = f(\Theta) + \epsilon \tag{6}$$

$$\epsilon \sim \mathcal{N}\left(0, \sigma^2(\Theta)\right) \tag{7}$$

where $f(\Theta)$ is the latent mean structure–property function and $\sigma^2(\Theta)$ represents input-dependent residual variability. A Gaussian process prior is assigned to the latent function,

$$f(\Theta) \sim \mathcal{GP}\left(m(\Theta), k(\Theta, \Theta')\right), \tag{8}$$

where $m(\Theta)$ is the prior mean function and $k(\Theta, \Theta')$ is the covariance kernel. In this work, the kernel is used only to provide a smooth probabilistic surrogate over the cone-angle design space; detailed kernel choices and training procedures are provided in the Supplementary Note.

In standard (homoscedastic) GPR the noise variance is assumed constant across all inputs:

$$\sigma^2(\Theta) = \sigma_n^2 \tag{9}$$

where the same residual variance is assigned to every cone-angle input. This is used here only as a baseline to examine the role of input-dependent noise modeling in the controlled one-dimensional analysis. In contrast, heteroscedastic GPR allows the noise variance to vary with the input. A polynomial-exponential form[29] is adopted as it guarantees positivity by construction, keeps the noise model interpretable through low-dimensional polynomial coefficients, and reduces to the homoscedastic case when the polynomial degree is zero. Following this, the input-dependent variance is modeled as

$$\sigma^2(\Theta) = \left[k_{al} \exp\left(\sum_{r=1}^{d} \sum_{j=1}^{p} a_{rj} \theta_j^r\right)\right]^2, \tag{10}$$

where $p = 3$ is the input dimension, $d$ is the polynomial degree, $k_{al} > 0$ is a scaling factor, and $a_{rj}$ are polynomial coefficients.

Given a training dataset $\mathcal{D} = \{(\Theta_i, y_i)\}_{i=1}^{N}$, the posterior predictive distribution at a test input $\Theta_*$ is expressed as $p(y_*|\Theta_*, \mathcal{D}) = \mathcal{N}(\mu_*, s_*^2)$. The predictive variance $s_*^2$ decomposes into two conceptually distinct contributions:

- Epistemic uncertainty (GP posterior variance of $f(\Theta_*)$): reflects the model's lack of knowledge due to sparse sampling of the design space. This component decreases as more training data are added and is expected to be large in sparsely sampled regions.
- Aleatoric uncertainty (the noise term $\sigma^2(\Theta_*)$): reflects the irreducible, realization-induced scatter that persists even with abundant data, because different GRF realizations under the same $\Theta$ produce genuinely different morphologies and stiffness values.

It is important to note that $\sigma^2(\Theta)$ is not a directly supervised empirical variance at each design point. Rather, it is inferred by the heteroscedastic GPR from the pattern of residuals across sparse one-realization-per-point observations. The controlled repeated-realization analyses presented in Fig. 2 serve only as diagnostic evidence that realization-induced scatter exists and is input-dependent — they are not used as variance labels during training.

Model assessment focuses on the reliability of the predictive distribution rather than mean prediction alone. Three uncertainty-oriented metrics are used: interval score (IS), miscalibration area, and negative log-likelihood (NLL). For a prediction interval $[L, U]$ with nominal coverage $1-\alpha$, the IS is defined as $IS_\alpha = (U-L) + \frac{2}{\alpha}(L-y)1(y<L) + \frac{2}{\alpha}(y-U\,)1(y>U)$. This penalizes both overly wide intervals and intervals that fail to contain the observation. The NLL is computed as $NLL = -\frac{1}{N}\sum_{i=1}^{N}\log p(y_i|\Theta_i)$, evaluating the full predictive distribution quality. The miscalibration area (Miscal. Area) is defined as the area between the expected and observed cumulative distributions in the calibration curve; smaller values indicate better calibration. These metrics are used in Section 4 to assess whether heteroscedastic GPR provides calibrated input-dependent predictive uncertainty. Training and test sets were constructed as described in Supplementary Note.

# 4. Results and Discussion

## 4.1. Realization-induced stiffness variability along a one-dimensional cone-angle

We first examine the stiffness response along a controlled one-dimensional cone-angle path, $\theta_1 \in [0°, 90°]$ with $\theta_2 = \theta_3 = 15°$ fixed, to characterize the structure of realization-induced scatter before introducing the surrogate model. Fig. 2 presents all nine independent components of the orthotropic stiffness tensor[1,7,21,22] along this path, where each point corresponds to a single GRF realization. Isolating $\theta_1$ in this way deliberately breaks the symmetry of the design space: the $\hat{\boldsymbol{e}}_2$- and $\hat{\boldsymbol{e}}_3$-directions experience a fixed, narrow cone throughout, while the $\hat{\boldsymbol{e}}_1$-direction undergoes a continuous structural transition from a tightly collimated to a broadly dispersed wave-vector distribution.

The mean stiffness trends reflect this asymmetry in a physically interpretable way, and they organize clearly according to whether the index 1 appears in the component subscript. Components involving the index 1 — the normal term $\mathrm{C}_{1111}$, the normal-shear coupling terms $\mathrm{C}_{1122}$ and $\mathrm{C}_{1133}$, and the shear terms $\mathrm{C}_{1212}$ and $\mathrm{C}_{1313}$ — all exhibit a non-monotonic, U-shaped response with a minimum at intermediate $\theta_1$. At small $\theta_1$, the wave-vector distribution is tightly concentrated near the $\hat{\boldsymbol{e}}_1$-axis, which promotes a strongly aligned, lamellar-like architecture with high load-bearing capacity in the $\hat{\boldsymbol{e}}_1$-direction and strong coupling to adjacent directions. As $\theta_1$ increases and the cone widens, this alignment is progressively disrupted, reducing the contribution of the $\hat{\boldsymbol{e}}_1$-direction to all associated stiffness components. However, as $\theta_1$ approaches $90°$, the wave vectors become distributed across the plane perpendicular to the $\hat{\boldsymbol{e}}_1$-axis, re-introducing structural connectivity in the $\hat{\boldsymbol{e}}_1$-direction and giving rise to the characteristic U-shaped recovery. Components that do not involve the index 1 — $\mathrm{C}_{2222}$, $\mathrm{C}_{3333}$, $\mathrm{C}_{2233}$, and $\mathrm{C}_{2323}$ — behave differently. Since $\theta_2$ and $\theta_3$ remain fixed at $15°$, these components are not subject to the same competing alignment effects, and they exhibit a mild inverse-U shape as $\theta_1$ increases.

Independent of the mean trend, the scatter around the predictive mean is markedly non-uniform along the path and differs across components. It is concentrated near $\theta_1 \approx 0°$, where the tight wave-vector cone makes the resulting morphology highly sensitive to realization-to-realization fluctuations in wave-vector directions and phase angles, and diminishes progressively toward $\theta_1 = 90°$, where the broad distribution causes individual realizations to converge toward a common structural character. These observations jointly establish that the mapping from cone-angle descriptor to effective stiffness is neither deterministic nor uniformly variable: each input corresponds to a distribution of stiffness values whose width depends on both the component and the location in the design space, motivating the heteroscedastic noise model formulated in Section 3.

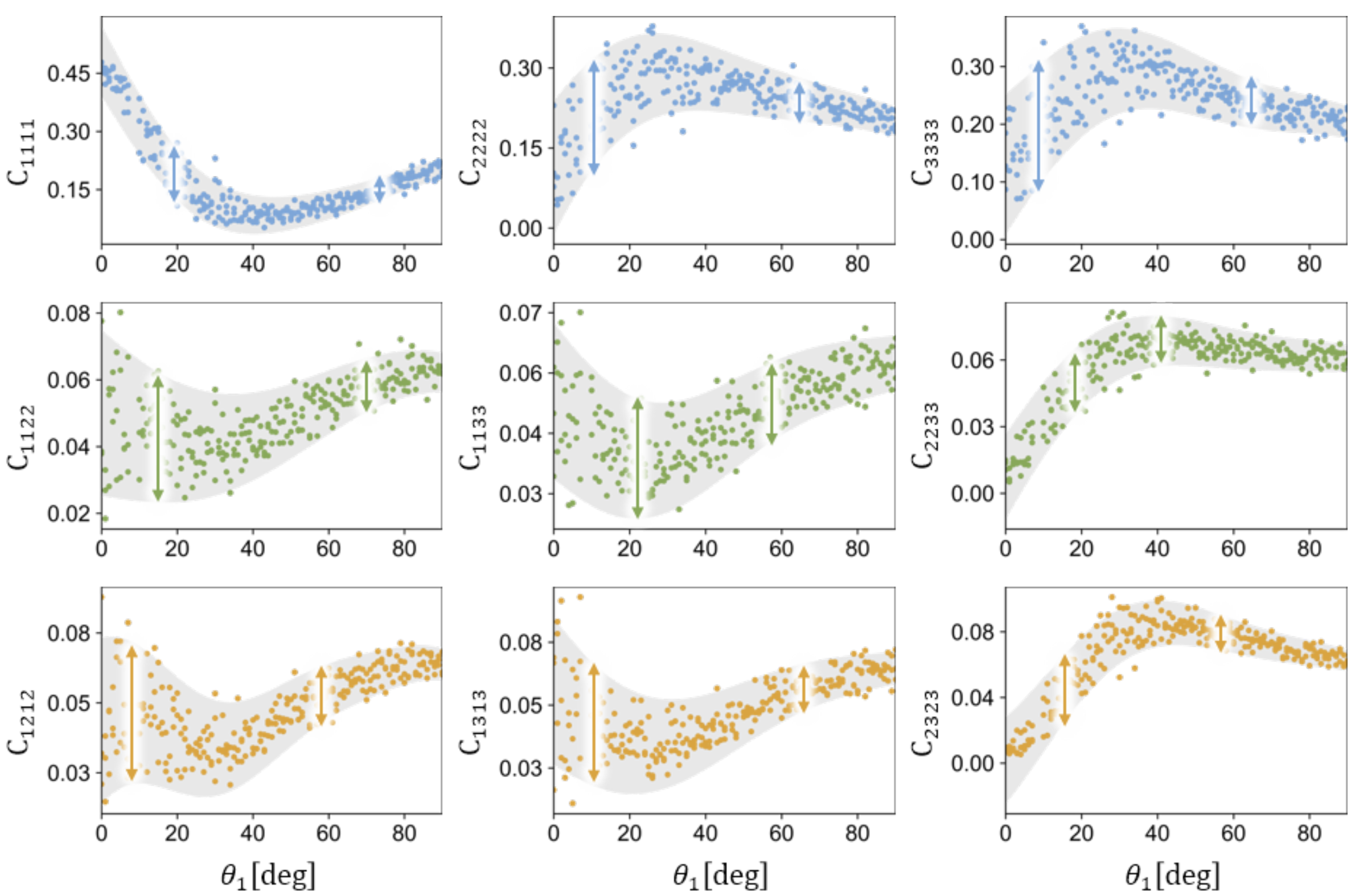

Fig. 2. Stiffness variability along a controlled one-dimensional cone-angle path. Nine independent stiffness components are evaluated as $\theta_1$ varies from $0°$ to $90°$, with $\theta_2 = \theta_3 = 15°$. The results show component-specific response trends and scatter patterns.

### 4.2. One-dimensional scenario: Homoscedastic vs. Heteroscedastic GPR

Before assessing whether a surrogate model trained on sparse one-realization-per-point data can recover this input-dependent scatter structure, we first explain why such data can contain information about the local noise level. The key lies in a distributional continuity argument: The GRF construction maps each cone-angle descriptor to a *distribution* of morphologies rather than to a single structure. The distributions $p(C_{ij}|\Theta)$ and $p(C_{ij}|\Theta + \delta\Theta)$ share overlapping support for any small $\delta\Theta$: if a stiffness value is achievable at $\Theta$, a morphologically similar realization at $\Theta + \delta\Theta$ can produce it as well. Conversely, if a gap were to appear between the two distributions as $\Theta$ is discontinuously varied, it would imply that some stiffness value achievable at $\Theta$ becomes entirely inaccessible at $\Theta + \delta\Theta$ — but this is contradicted by the fact that the same $\Theta + \delta\Theta$ can generate arbitrarily varied morphologies through different random seeds. This suggests that abrupt distributional gaps are unlikely under the GRF construction. It follows that sparse training observations drawn from neighboring design points — each a single realization at a distinct $\Theta$ — constitute approximate samples from $p(C_{ij}|\Theta_0)$ in a neighborhood of any target $\Theta_0$. The spatial residual pattern they include relative to the GP mean encodes information about the local noise level $\sigma(\Theta_0)$, and heteroscedastic GPR exploits this by maximizing the marginal likelihood over both the kernel hyperparameters and the noise function jointly — extracting $\sigma(\Theta)$ from residual structure without requiring explicit variance labels at any design point. Fig. 3(a) and (b) compare homoscedastic and heteroscedastic GPR predictions along the one-dimensional path for three representative components, and Fig. 3(c) isolates the uncertainty contributions separately: the solid lines track the inferred aleatoric noise, and the dashed lines track the epistemic GP posterior variance, for each model.

Both models recover the mean stiffness trend with comparable fidelity, as is expected — the mean prediction is governed primarily by the GP kernel and is largely insensitive to how the noise is modeled. The substantive difference lies in the structure of the predictive uncertainty. The homoscedastic model is constrained to assign a single noise level across the entire path; in effect, it averages the scatter that is genuinely concentrated near small $\theta_1$ with the near-deterministic behavior at large $\theta_1$. This results in a predictive band that is overconfident in the low-scatter regime and insufficiently expressive in the high-scatter regime. The heteroscedastic model, by contrast, learns a noise function that decreases with $\theta_1$ — without being supplied any explicit variance labels — and concentrates the aleatoric uncertainty budget where the underlying data are genuinely variable.

A further consequence of the correct noise model, visible in the dashed lines of the right column, concerns the epistemic uncertainty. At large $\theta_1$, where the data are nearly noiseless, the epistemic variance of the heteroscedastic model is meaningfully lower than that of the homoscedastic model. This occurs because, under maximum marginal likelihood training, the homoscedastic model must explain the high scatter at small $\theta_1$ with a large noise parameter, which then inflates the effective noise attributed to the low-scatter region as well. The heteroscedastic model does not face this trade-off: by accurately localizing the aleatoric contribution, it allows the GP kernel to fit the low-scatter region more tightly, yielding lower posterior variance there. In practical terms, this means that the heteroscedastic model extracts more information from the same training set by allocating the signal-to-noise ratio appropriately across the design path.

The quantitative improvement is summarized in Table 1. Across all nine components, the heteroscedastic model achieves lower NLL, lower IS, and lower Miscal. Area than the homoscedastic baseline, while the MAE remains essentially unchanged, confirming that the benefit is in calibration rather than mean accuracy. The fitted noise functions in Table 1 are all of the form $\sigma(\theta_1) = k_{al} \exp(-a\theta_1)$ with $a > 0$, a first-order polynomial in the exponent. The selection of polynomial degree $d = 1$ was based on expected log predictive density ($ELPD_{approx}$)[32], which estimates the Leave-One-Out (LOO) Cross-Validation[33] criterion without requiring model retraining at each held-out point, evaluated over held-out data (Supplementary Note); but the fact that a linear exponent suffices has a physical interpretation worth noting: it implies that the logarithm of the noise variance decreases linearly with $\theta_1$, or equivalently, that each incremental widening of the cone angle reduces scatter by a constant multiplicative factor. This is consistent with the nature of a GRF whose admissible wave-vector set expands smoothly and uniformly as $\theta_1$ increases — there is no abrupt structural transition, and accordingly, the scatter does not exhibit any threshold or saturation behavior that would require higher-order terms. As will be shown in Section 4.3, the same degree selection holds across the full three-dimensional parameter space, suggesting that this linear-in-exponent structure is not particular to the one-dimensional path examined here but reflects a more general property of the spinodoid uncertainty landscape.

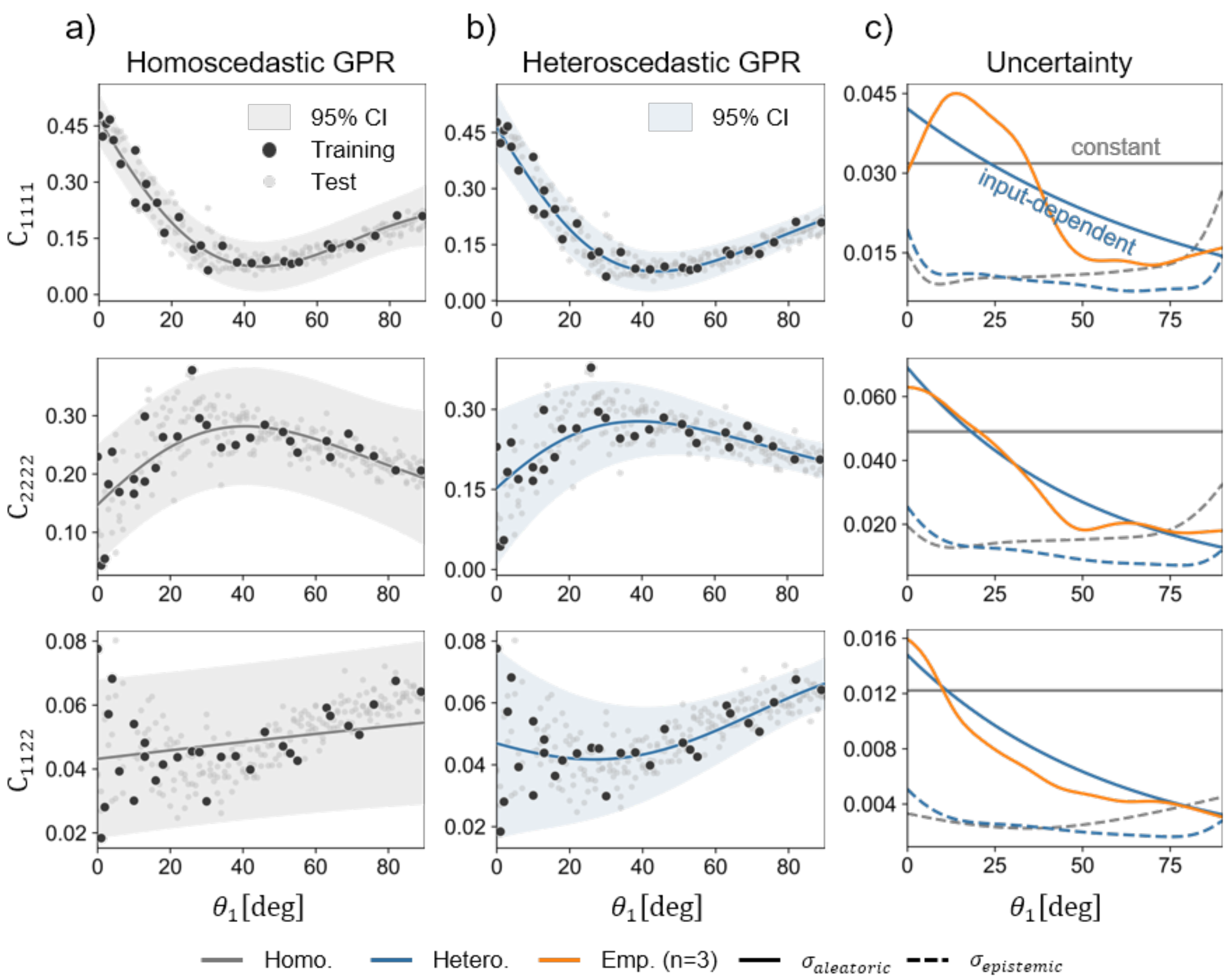

Fig. 3. One-dimensional comparison of homoscedastic and heteroscedastic GPR. Homoscedastic GPR provides a constant noise, whereas heteroscedastic GPR infers a $\theta_1$-dependent predictive uncertainty along the controlled cone-angle path. Emp. (n=3): empirical standard deviation from three GRF realizations per point, locally smoothed; provided as a small-sample diagnostic reference.

Table 1. Controlled one-dimensional comparison of homoscedastic and heteroscedastic GPR models for component-wise uncertainty estimation. The controlled path is defined by $\theta_1 \in [0°, 90°]$ with $\theta_2 = \theta_3 = 15°$. Miscalibration area (Miscal. Area), negative log-likelihood (NLL), mean absolute error (MAE), and interval score (IS) are reported as homoscedastic/heteroscedastic values, respectively. Downward arrows indicate that lower values are preferred.

| Component | Homo. | Hetero. | Miscal. Area ↓ | NLL ↓ | MAE ↓ | IS ↓ |
|---|---|---|---|---|---|---|
| $C_{1111}$ | 0.0318 | 0.0422exp(–1.0766$\theta_1$) | 0.0441/0.0102 | –2.1550/–2.3096 | 0.0204/0.0202 | 0.1481/0.1244 |
| $C_{2222}$ | 0.0491 | 0.0691exp(–1.6943$\theta_1$) | 0.0656/0.0224 | –1.8100/–2.0975 | 0.0273/0.0272 | 0.2047/0.1478 |
| $C_{3333}$ | 0.0527 | 0.0689exp(–1.4608$\theta_1$) | 0.0867/0.0298 | –1.8171/–2.1231 | 0.0253/0.0250 | 0.2106/0.1471 |
| $C_{1122}$ | 0.0122 | 0.0148exp(–1.5233$\theta_1$) | 0.0574/0.0159 | –3.2241/–3.6434 | 0.0072/0.0054 | 0.0500/0.0329 |
| $C_{1133}$ | 0.0103 | 0.0116exp(–1.2291$\theta_1$) | 0.0293/0.0456 | –3.3093/–3.6844 | 0.0070/0.0051 | 0.0428/0.0295 |
| $C_{2233}$ | 0.0071 | 0.0095exp(–1.1344$\theta_1$) | 0.0633/0.0628 | –3.6643/–3.7255 | 0.0047/0.0048 | 0.0318/0.0294 |
| $C_{1212}$ | 0.0136 | 0.0182exp(–1.8424$\theta_1$) | 0.0912/0.0121 | –3.1854/–3.5659 | 0.0062/0.0061 | 0.0554/0.0342 |
| $C_{1313}$ | 0.0170 | 0.0201exp(–1.7731$\theta_1$) | 0.0517/0.0461 | –2.9057/–3.5496 | 0.0099/0.0058 | 0.0694/0.0390 |
| $C_{2323}$ | 0.0074 | 0.0111exp(–1.6543$\theta_1$) | 0.0586/0.0215 | –3.3544/–3.4676 | 0.0057/0.0057 | 0.0425/0.0399 |

### 4.3. Three-dimensional scenario: Extension and consistency with the one-dimensional analysis

We now extend the heteroscedastic GPR framework to the full three-dimensional cone-angle design space, where $\theta_1$, $\theta_2$, and $\theta_3$ vary independently over $[0°, 90°]^3$. Fig. 4 presents two complementary diagnostics for $\mathrm{C}_{1111}$, $\mathrm{C}_{2222}$, and $\mathrm{C}_{1122}$, comparing the homoscedastic and heteroscedastic models. Fig. 4(a) shows calibration curves for each model, where the empirical cumulative distribution function (ECDF) of the normalized residuals, $(y_{true} - y_{pred})/\sigma_{pred}$, is plotted against the CDF of a standard Gaussian[34]. For a perfectly calibrated model, the normalized residuals follow a standard Gaussian and the ECDF lies on the diagonal; the miscalibration area quantifies the departure from this ideal one. The heteroscedastic curves lie close to the diagonal for all three components, while the homoscedastic curves exhibit an S-shaped departure — the ECDF falls below the diagonal through the central range and rises above it at the tails. This S-shape is a characteristic signature of a constant-noise model applied to heteroscedastic data: by assigning a single global $\sigma$ that averages over regions of very different scatter, the homoscedastic model produces normalized residuals whose distribution is over-dispersed relative to a standard Gaussian, with heavier tails and a flatter center than expected. Fig. 4(b) reinforces this interpretation by plotting the estimated noise $\sigma(\Theta)$ against the actual absolute residual $|y_{true} - y_{pred}|$ for each test point. The homoscedastic points collapse onto a horizontal strip — a single constant $\sigma$ value for all test points — demonstrating that the model has no capacity to anticipate where prediction errors will be large. The heteroscedastic points span a range of $\sigma(\Theta)$ values and exhibit a positive correlation with the actual residuals, confirming that the inferred noise function encodes genuine spatial structure in the design space rather than a uniform average. The improvements in Miscal. Area, NLL, and IS reported in Table 2 reflect these two benefits.

Table 2 also reveals that the polynomial degree $d = 1$ is selected for all nine components by $ELPD_{LOO}$, confirming that the linear-in-exponent noise structure identified in Section 4.2 is not particular to the one-dimensional path but holds throughout the full parameter space. The fitted coefficients carry a physically transparent interpretation: each component is most sensitive, in terms of noise decay, to the cone angles governing its associated load paths. $\mathrm{C}_{1212}$ exhibits the largest coefficients on $\theta_1(-1.55)$ and $\theta_2(-1.82)$, reflecting that in-plane shear draws on structural connectivity in both the $\hat{\boldsymbol{e}}_1$- and $\hat{\boldsymbol{e}}_2$-directions, and its scatter is suppressed most effectively when both cones are widened. By contrast, the $\theta_2$ coefficient of $\mathrm{C}_{1313}(-0.13)$ is substantially smaller than its $\theta_1(-1.15)$ and $\theta_3(-1.37)$ counterparts, consistent with 1-3 shear being governed primarily by the $\hat{\boldsymbol{e}}_1$- and $\hat{\boldsymbol{e}}_3$-directional wave-vector content and relatively less sensitive to the breadth of the $\theta_2$ cone. A similar asymmetry appears in the coupling components: the $\theta_2$ coefficient of $\mathrm{C}_{1133}(-0.28)$ is markedly smaller than its $\theta_1(-0.84)$ and $\theta_3(-1.18)$ values, while for $\mathrm{C}_{2233}$ the $\theta_1$ coefficient $(-0.42)$ is notably smaller than those on $\theta_2(-1.21)$ and $\theta_3(-1.31)$. In both cases, the relatively weak coefficient corresponds to the cone angle of the axis not directly involved in the coupling — the $\hat{\boldsymbol{e}}_2$-direction for $\mathrm{C}_{1133}$ and the $\hat{\boldsymbol{e}}_1$-direction for $\mathrm{C}_{2233}$. Rather than asserting that these coefficients are negligible, the more precise observation is that the noise model recovers a directional hierarchy that is physically consistent with the load-path structure of each tensor component, with the dominant noise-suppression axes aligning with the mechanically active directions.

Fig. 5 maps the predicted mean and noise function onto the $\theta_1$– $\theta_2$ plane at $\theta_3 = 15°$. The mean fields recover the component-specific trends established in Section 4.1: $\mathrm{C}_{1111}$ shows a non-monotonic contour structure reflecting the U-shaped response along $\theta_1$, $\mathrm{C}_{2222}$ exhibits the symmetric pattern along $\theta_2$, and $\mathrm{C}_{1122}$ decreases more gradually with both cone angles. The uncertainty maps reveal a distinct spatial structure: the iso-uncertainty contours run diagonally across the $\theta_1$– $\theta_2$ plane, with highest uncertainty in the low-$\theta_1$, low-$\theta_2$ corner. This orientation follows directly from the additively separable log-noise function in Table 2 — because $\log \sigma(\Theta) = \log k_{al} - a_1\theta_1 - a_2\theta_2 - a_3\theta_3$, the iso-uncertainty contours at fixed $\theta_3$ are straight lines with slope $-a_1/a_2$. For $C_{1111}$, where $a_2(0.90)$ exceeds $a_1(0.54)$, the contours tilt more steeply toward the $\theta_2$-axis, indicating that widening $\theta_2$ suppresses scatter in $C_{1111}$ more effectively than widening $\theta_1$. For $C_{2222}$, $a_1(0.83)$ and $a_2(0.95)$ are nearly equal, yielding near-symmetric contours. The choice of $\theta_3 = 15°$ for the displayed slice is representative: additive separability implies that varying $\theta_3$ shifts all contours uniformly in absolute level — by an amount proportional to $a_3 \cdot \Delta\theta_3$ — while preserving their orientation. The $\theta_3$ coefficients in Table 2 range from $-0.53(C_{2222})$ to $-1.63(C_{2323})$, so the magnitude of this shift is component-specific and can be substantial. Mean and uncertainty maps at additional $\theta_3$ values are provided in Supplementary Note, confirming that the diagonal contour structure is preserved across the full range of the third cone angle. From a design perspective, the mean and uncertainty maps together provide information inaccessible from a deterministic surrogate: a designer targeting a specific stiffness can identify not only which cone-angle combination achieves the desired mean, but which choice carries the lowest aleatoric uncertainty and therefore the lowest risk of the fabricated sample deviating from the intended response.

Finally, Fig. 6 cross-validates the 3D noise model against the controlled one-dimensional analysis by evaluating the 3D noise functions along the slice $\theta_2 = \theta_3 = 15°$, corresponding precisely to the path used in Sections 4.1 and 4.2. Both the 1D-fitted curves and the 3D curves evaluated on this slice decrease monotonically with $\theta_1$ for all nine components, and the component ordering in noise magnitude is preserved between the two models. A systematic quantitative difference is nevertheless present: the 1D curves decay more steeply with $\theta_1$ than the 3D curves, a pattern confirmed directly by comparing the $\theta_1$ coefficients in Tables 1 and 2. For $C_{1212}$, the 1D coefficient is $-1.84$ while the 3D coefficient is $-1.55$; for $C_{2233}$ the discrepancy is larger ($-1.13$ vs. $-0.42$). This attenuation reflects the difference between marginal and conditional noise attribution. The 1D model, trained exclusively along the path $\theta_2 = \theta_3 = 15°$, attributes all observed noise variation to the $\theta_1$ coefficient. The 3D model, trained where all three cone angles vary simultaneously, distributes the noise-reduction effect across $a_1$, $a_2$, and $a_3$: variation that the 1D model assigns entirely to $\theta_1$ is, in the broader dataset, partially correlated with variation in $\theta_2$ and $\theta_3$, and is absorbed by their respective coefficients. The 1D coefficient therefore reflects the total marginal noise reduction along the controlled path, while the 3D $\theta_1$ coefficient reflects only the partial contribution after conditioning on the other two angles. That the monotonic trend is preserved across all nine components despite this redistribution — and across two models trained on entirely different datasets — constitutes independent evidence that the concentration of stiffness scatter at small cone angles is a structural property of the spinodoid design space, not an artifact of any particular training set or model parametrization.

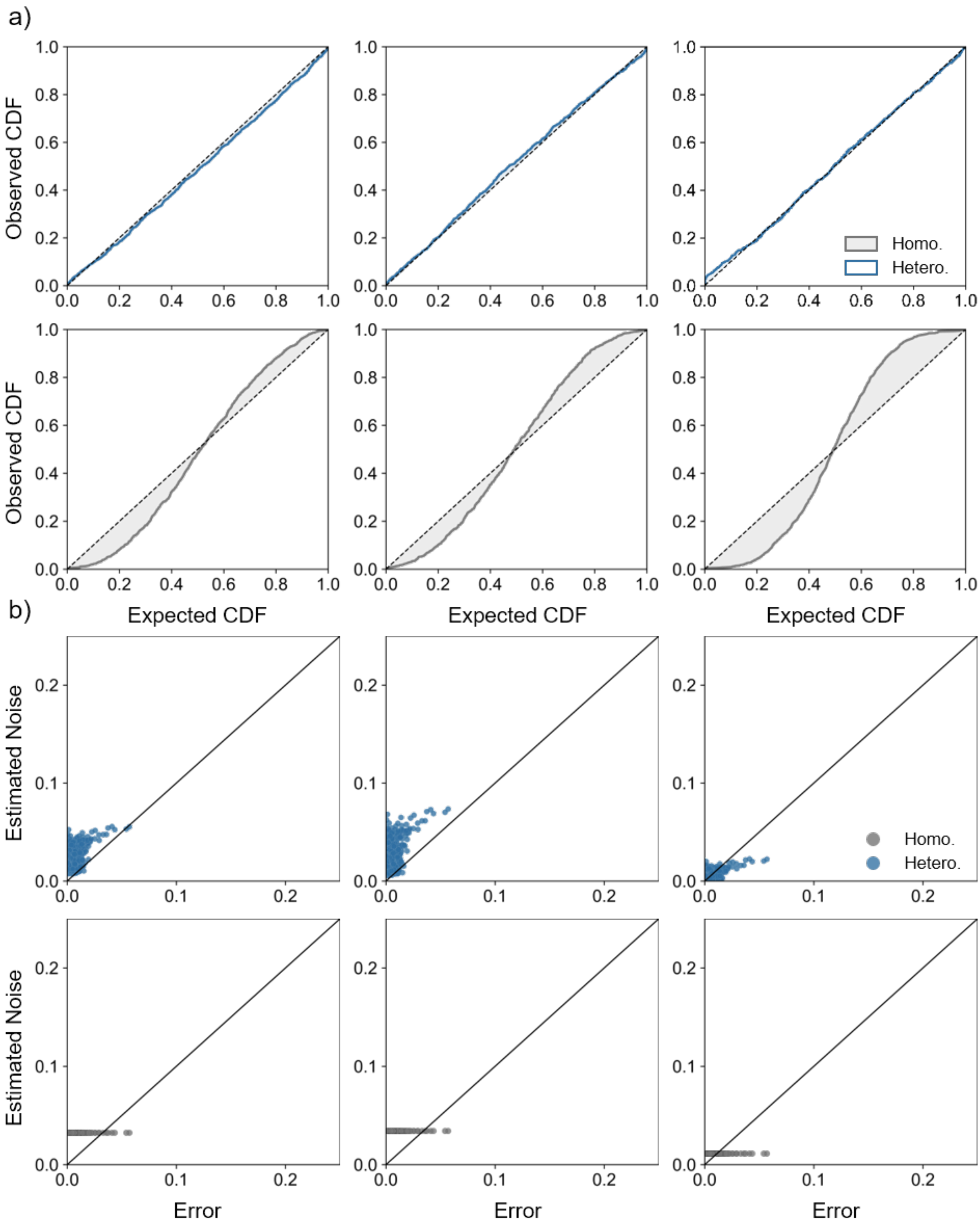

Fig. 4. Three-dimensional heteroscedastic GPR prediction and calibration. (a) Calibration and (b) noise-error plots for selected stiffness components in the full cone-angle design space. The results assess both mean prediction and predictive uncertainty calibration.

Table 2. Three-dimensional heteroscedastic GPR noise models and relative improvements in uncertainty metrics over the full cone-angle design space $\Theta = (\theta_1, \theta_2, \theta_3)$.

| Component | Hetero. | Miscal. Area | NLL | MAE | IS |
|---|---|---|---|---|---|
| $C_{1111}$ | $0.0489\exp(-0.5433\theta_1-0.9026\theta_2-0.5800\theta_3)$ | 87.65% | 63.60% | 74.89% | 78.95% |
| $C_{2222}$ | $0.0653\exp(-0.8342\theta_1-0.9486\theta_2-0.5315\theta_3)$ | 82.07% | 8.03% | –1.32% | 22.80% |
| $C_{3333}$ | $0.0774\exp(-1.3217\theta_1-0.8707\theta_2-0.8252\theta_3)$ | 70.38% | 4.18% | –4.79% | 21.92% |
| $C_{1122}$ | $0.0217\exp(-1.1329\theta_1-1.6754\theta_2-1.1793\theta_3)$ | 92.68% | 8.19% | –4.13% | 37.24% |
| $C_{1133}$ | $0.0118\exp(-0.8376\theta_1-0.2750\theta_2-1.1821\theta_3)$ | 76.51% | 4.41% | 4.15% | 17.98% |
| $C_{2233}$ | $0.0158\exp(-0.4186\theta_1-1.2070\theta_2-1.3084\theta_3)$ | 30.83% | 0.42% | –5.27% | 15.39% |
| $C_{1212}$ | $0.0321\exp(-1.5504\theta_1-1.8249\theta_2-0.8442\theta_3)$ | 69.73% | 15.47% | –1.66% | 51.34% |
| $C_{1313}$ | $0.0174\exp(-1.1529\theta_1-0.1277\theta_2-1.3678\theta_3)$ | 56.76% | 7.12% | 9.48% | 26.80% |
| $C_{2323}$ | $0.0209\exp(-0.5821\theta_1-1.4535\theta_2-1.6253\theta_3)$ | 75.34% | 2.50% | –4.84% | 21.33% |

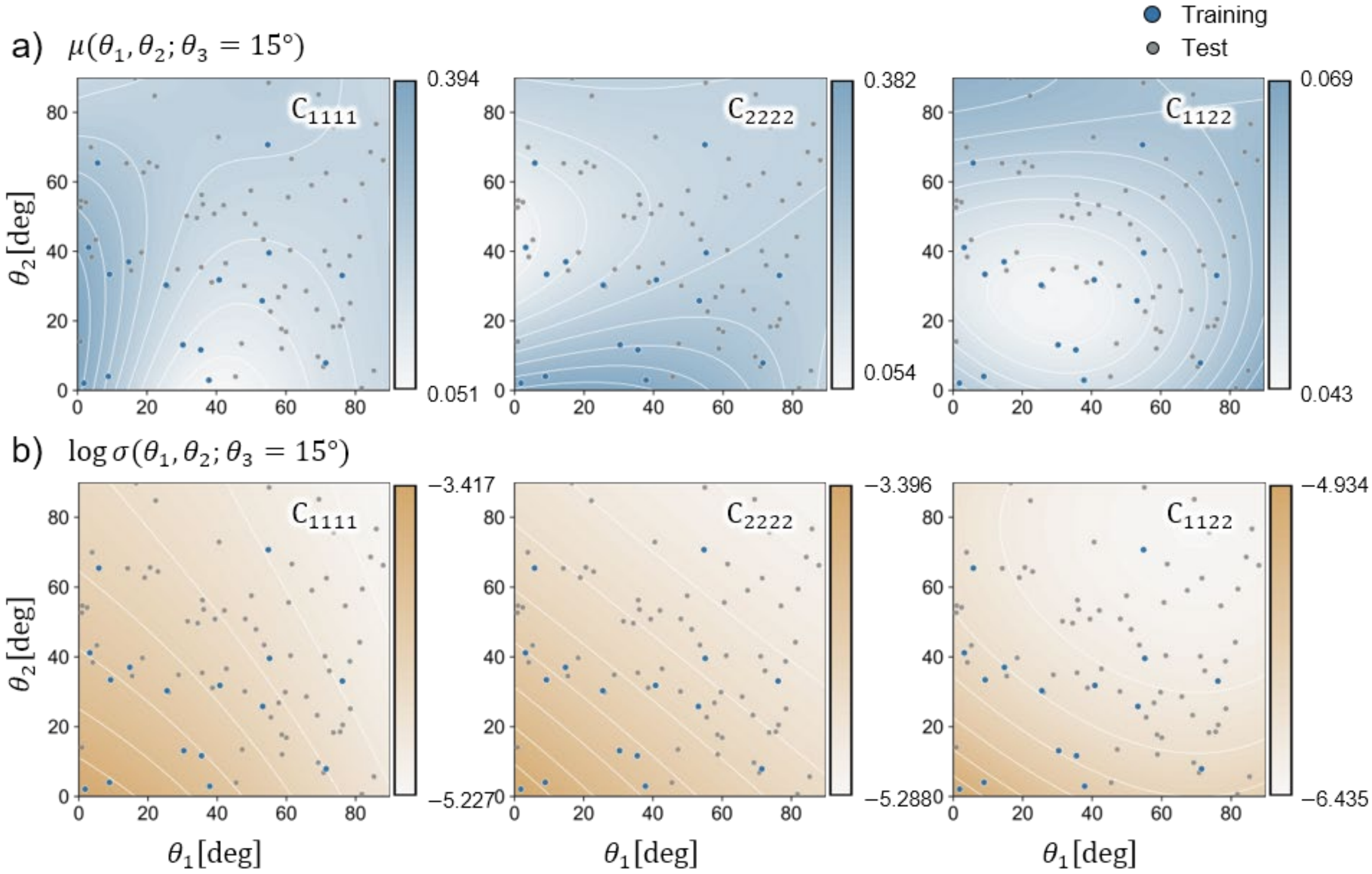

Fig. 5. Heteroscedastic GPR-predicted (a) mean and (b) uncertainty maps for selected stiffness components. The uncertainty maps provide a design space interpretation of predictive reliability.

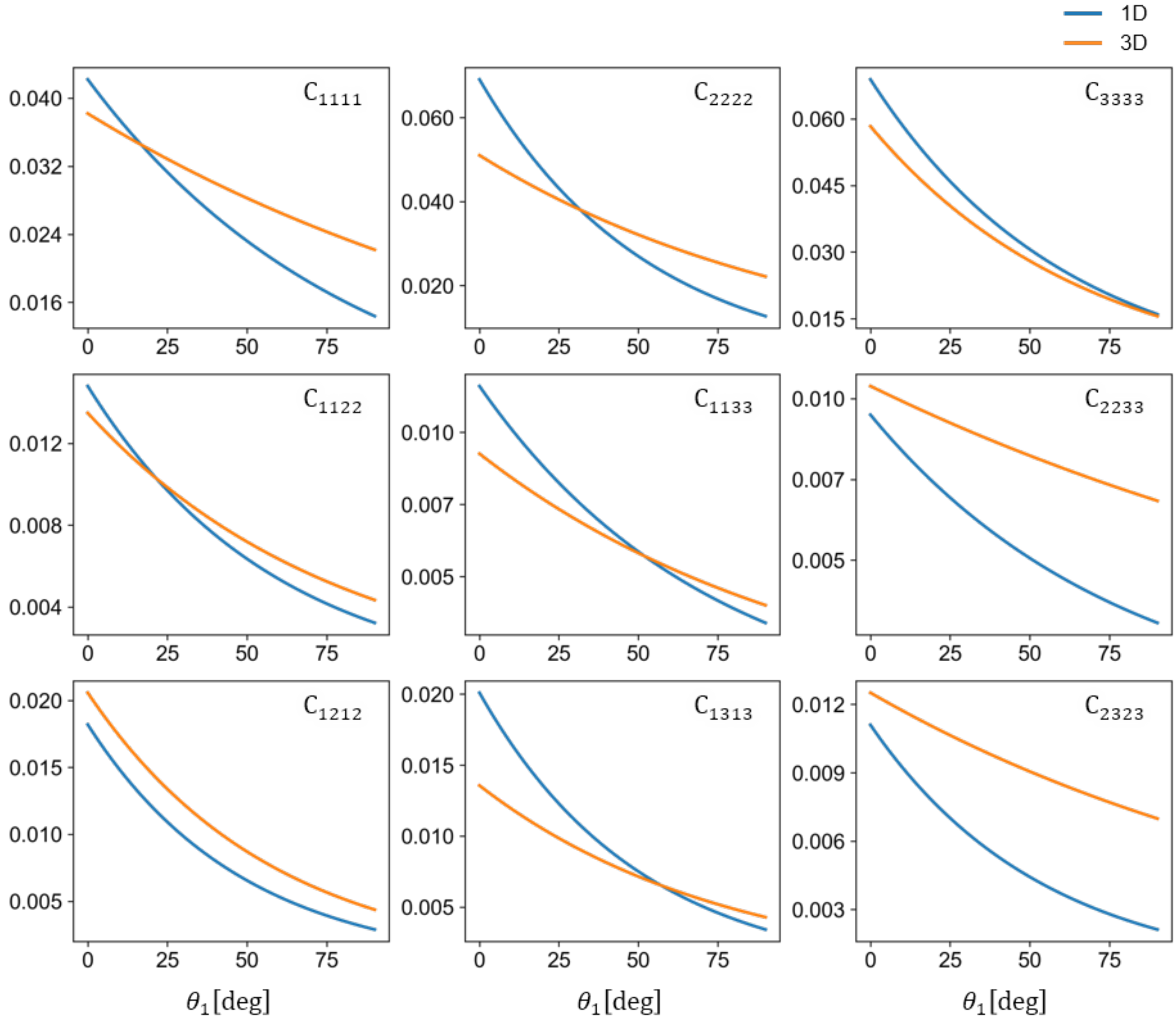

Fig. 6. Trend-level consistency between 1D and 3D heteroscedastic noise trends.

### 4.4 Application to RBDO

This section demonstrates DDO and RBDO[35,36] for spinodoid metamaterials using uncertainty-aware surrogate models. The design variable is the cone-angle descriptor $\Theta = (\theta_1, \theta_2, \theta_3) \in [0°, 90°]^3$, and the design objective is to achieve a prescribed target value of the axial Young's modulus, $E_1 = 0.3$. At the same time, the stiffness ratios $E_2/E_1$ and $E_3/E_1$ are assumed to be sufficiently accurate over the considered design domain. For each modulus component $E_i$, the corresponding surrogate model provides the prediction mean $\hat{\mu}(\theta)$ and the morphology-induced aleatoric uncertainty (i.e., standard deviation) $\hat{\sigma}_i$, where $i$ = 1, 2, 3. Accordingly, the stochastic stiffness response used in the RBDO formulations can be expressed as

$$\tilde{E}_i(\Theta) = \hat{\mu}_i(\theta) + \hat{\sigma}_i(\theta)\xi_i, \qquad i = 1,2,3, \tag{11}$$

where $\xi = (\xi_1, \xi_2, \xi_3)^T$, and $\xi_i$ is a standardized random variable representing stochastic variation induced by different GRF morphology realizations. Under this common problem definition, one DDO and two RBDO formulations can be derived.

First, the DDO problem is formulated based solely on the predicted mean responses:

$$\Theta^*_{DDO} = \arg\min_{\Theta \in \Omega_\Theta} \frac{|\hat{\mu}_1(\Theta) - E_1|}{E_1} \tag{12}$$

$$\text{s.t.} \quad g(\Theta) \leq 0 \tag{13}$$

where

$$g(\Theta) = \max\left[\kappa - \frac{\hat{\mu}_2(\Theta)}{\hat{\mu}_1(\Theta)}, \kappa - \frac{\hat{\mu}_3(\Theta)}{\hat{\mu}_1(\Theta)}\right]. \tag{14}$$

Here, $g(\Theta) \leq 0$ indicates that both stiffness-ratio constraints are satisfied based on the predicted mean responses (i.e., $\kappa = 0.5$).

Second, the RBDO problem is formulated based on the predicted mean responses, while incorporating morphology-induced uncertainty into each stiffness component:

$$\Theta^*_{RBDO} = \arg\min_{\Theta \in \Omega_\Theta} \frac{\left|\tilde{E}_1(\Theta) - E_1\right|}{E_1} \tag{15}$$

$$\text{s.t.} \quad \Pr[G(\Theta, \xi) > 0] \le p_f^{target} \tag{16}$$

where

$$G(\Theta, \xi) = \max\left[\kappa - \frac{\tilde{E}_2(\Theta)}{\tilde{E}_1(\Theta)}, \kappa - \frac{\tilde{E}_3(\Theta)}{\tilde{E}_1(\Theta)}\right] \tag{17}$$

Here, $G(\Theta, \xi) \le 0$ indicates that both stiffness-ratio constraints are satisfied based on the stochastic stiffness responses, whereas $G(\Theta, \xi) > 0$ represents failure. Therefore, $\Pr[G(\mathit{\Theta}, \xi) > 0]$ denotes the probability, which is required to be no greater than the target failure probability, $p_f^{target}$ = 0.01. In addition, two surrogate-based RBDO cases are performed. RBDO-HOMO uses homoscedastic GPR with spatially uniform predictive uncertainty, whereas RBDO-HETERO uses heteroscedastic GPR with input-dependent predictive uncertainty. For each case, the failure probability is estimated by Monte Carlo simulations (MCS) using $10^6$ samples from the corresponding surrogate predictive distribution. Note that all other RBDO settings are kept identical, so any difference between the two cases arises soley from the representation of uncertainty.

Table 3 summarizes the optimization results. The DDO solution almost exactly matches the target stiffness because it only optimizes the deterministic mean response. By contrast, the two RBDO solutions sacrifice target-matching accuracy to satisfy the prescribed reliability requirement. As a result, both RBDO-HOMO and RBDO-HETERO yield more conservative designs, reducing their own-model failure probabilities below the target value, with $p_f$ = 0.0089 and 0.0087, respectively. However, as shown in Fig. 2, the uncertainty of the output responses varies depending on the input, so each design's own-model estimate may not reflect its true reliability. To assess this directly and independently of either surrogate model, the failure probability at each optimized design was additionally estimated from 1,000 realizations using computational homogenization, yielding the 95% Clopper-Pearson confidence interval[37] for the empirical failure probability, denoted $p_f^{CI}$. This direct validation confirms that the deterministic optimum is highly vulnerable to constraint violation, with $p_f^{CI}$ = [0.5258, 0.6072] lying far above the target failure probability. More importantly, depite satisfying its own reliability target, the RBDO-HOMO design yields $p_f^{CI}$ = [0.0272, 0.0542], entirely above the target failure probability. These results demonstrate the importance of incorporating uncertainty into reliable design. In particular, assuming homoscedastic uncertainty can result in an unreliable design when the actual response uncertainty varies across the input space. By directly accounting for this input-dependent uncertainty, RBDO-HETERO provides a more appropriate framework for reliability-aware design optimization.

In summary, these results highlight the necessity of heteroscedastic RBDO for reliable inverse design of spinodoid metamaterials. The comparison shows that a deterministic optimum can be misleading when morphology-induced variability is ignored, and that a homoscedastic uncertainty assumption may fail to properly represent the actual failure probability. More importantly, accounting for input-dependent predictive uncertainty is essential for obtaining reliable designs when morphology-induced variability is present. Thus, the heteroscedastic RBDO formulation offers a more reliable and physically consistent design framework by reflecting the spatially varying uncertainty inherent in the spinodoid structure–property relationship.

Table 3. Comparison of deterministic and reliability-based inverse design results.

| Method | $\Theta$ [deg] | Objective* | Mean ratios ($E_2/E_1$, $E_3/E_1$) | $p_f^{\text{Own}}$ ** | $p_f^{CI}$*** | Feasibility*** |
|---|---|---|---|---|---|---|
| DDO | [8, 21, 22] | $9.4 \times 10^{-15}$ | (0.58, 0.56) | 0.49 (> 0.01) | [0.5258, 0.6072] (>0.01) | Violated |
| RBDO-HOMO | [16, 23, 25] | $1.8 \times 10^{-1}$ | (0.78, 0.80) | $8.9 \times 10^{-3}$ ($\leq$ 0.01) | [0.0272, 0.0542] (>0.01) | Violated |
| RBDO-HETERO | [18, 54, 51] | $2.7 \times 10^{-1}$ | (0.73, 0.73) | $8.7 \times 10^{-3}$ ($\leq$ 0.01) | [0.0000, 0.0058] ($\leq$0.01) | Satisfied |

* The optimized $E_1$ values are 0.30, 0.25, and 0.22 for DDO, RBDO-HOMO and RBDO-HETERO, respectively.

** $p_f^{\text{Own}}$ is evaluated at each optimized design using the corresponding uncertainty model.

*** $p_f^{CI}$ denotes the 95% confidence interval of the empirical failure probability estimated from 1,000 realizations at each optimized design (Clopper-Pearson confidence interval[37]). The feasibility of each optimized design is therefore assessed based on $p_f^{CI}$.

## 5. Conclusion

Spinodoid metamaterials generated by Gaussian random fields occupy a distinctive position in the landscape of architected materials: their mechanical properties are not determined by a single realization but by an ensemble of morphologies, all consistent with the same cone-angle descriptor. This study has shown that this stochasticity should not simply be averaged out; it is a structured, input-dependent phenomenon whose variation across the design space carries physically meaningful information and has measurable consequences for design reliability.

The one-dimensional analysis established that the nine independent stiffness components respond to varying $\theta_1$ in ways governed by the directional structure of their load paths. Components involving the index 1 exhibit non-monotonic, U-shaped trends arising from the competition between preferential alignment along the 1-direction and wave-vector redistribution at large cone angles, whereas components not involving index 1 exhibit an inverse-U trend along the asymmetric path $\theta_2 = \theta_3 = 15°$. Scatter is concentrated at small cone angles for all components, and its magnitude follows the same directional logic. The three-dimensional analysis showed that heteroscedastic GPR recovers this input-dependent noise structure from sparse one-realization-per-point observations, without access to explicit variance labels. Calibration diagnostics confirm that the learned noise function reflects genuine spatial structure in the design space rather than a uniform average, and the fitted polynomial degree $d = 1$, selected consistently across all nine components in both the 1D and 3D settings, implies that each incremental cone-angle widening reduces scatter by a constant multiplicative factor, with no threshold or saturation behavior. The 1D–3D consistency analysis further showed that this property is not an artifact of any particular training set, but a structural feature of spinodoid uncertainty.

The practical utility of this framework was demonstrated through DDO and RBDO. A deterministic optimum achieves close agreement with the target stiffness on the basis of mean predictions alone, yet is shown to be highly susceptible to stiffness-ratio constraint violation once morphology-induced variability is accounted for. An RBDO formulation based on homoscedastic uncertainty likewise fails to meet the prescribed reliability requirement when evaluated against the input-dependent scatter estimated by the heteroscedastic surrogate. In the present surrogate-based evaluation, only the heteroscedastic RBDO formulation, which directly incorporates the spatial variation of aleatoric uncertainty into the constraint probability, satisfies the reliability target. These results underscore a broader point: when multiple cone-angle combinations achieve nominally equivalent mean responses, their aleatoric uncertainty levels can differ substantially, so that a deterministically adequate design may carry very different fabrication risk depending on its location in the design space.

Several extensions merit consideration. The present framework treats aleatoric uncertainty as arising solely from GRF realization variability at fixed nominal cone-angle parameters; in practice, fabrication processes introduce additional scatter through geometric tolerances on the realized cone angles themselves. Incorporating this manufacturing-induced uncertainty as a separable second source of aleatoric noise would bring the framework closer to the conditions of physical production. A second direction concerns the independent treatment of the nine stiffness components: training a separate surrogate for each component neglects the physical correlations among them, and extending to multi-output GPR through, for instance, a linear model of coregionalization would enable joint probabilistic predictions of the full stiffness tensor. This is particularly relevant for RBDO formulations whose constraints involve ratios of multiple components, where the joint distribution rather than marginal uncertainties governs the failure probability. Finally, extending the uncertainty-aware surrogate to nonlinear mechanical responses and integrating it with active learning schemes that concentrate additional sampling in high-scatter regions represent natural continuations of the present work.

## Acknowledgements

This work supported by ~

*Supplementary Note*

# Uncertainty-Aware Structure–Property Mapping of Spinodoid Metamaterials *via* Heteroscedastic Gaussian Process Regression

Minwoo Park[a]†, Junseo Park[a]†, Mingyu Lee[a]†, Hugon Lee[a], Hanbin Cho[a], Ikjin Lee[abc]*, and Seunghwa Ryu[abc]*

[a]Department of Mechanical Engineering, Korea Advanced Institute of Science and Technology (KAIST), Daejeon 34141, Republic of Korea

[b]KAIST InnoCORE PRISM-AI Center, Korea Advanced Institute of Science and Technology (KAIST), Daejeon 34141, Republic of Korea

[c]Department of AX, Korea Advanced Institute of Science and Technology (KAIST), Daejeon 34141, Republic of Korea

†These authors have contributed equally: Minwoo Park, Junseo Park, and Mingyu Lee

*Corresponding author. E-mail: ryush@kaist.ac.kr, and ikjin.lee@kaist.ac.kr

# S1. Details of Heteroscedastic Gaussian Process Surrogate Model

## S1.1 Kernel Function and Hyperparameter Optimization

Each spinodoid metamaterial is parameterized by the design descriptor $\Theta = (\theta_1, \theta_2, \theta_3) \in \mathbb{R}^3$. For each independent stiffness component $C_k(\Theta, \omega)$, a Gaussian process (GP) surrogate is constructed with a radial basis function (RBF) kernel with automatic relevance determination (ARD):

$$k(\theta, \theta') = \sigma_f^2 \exp\left(-\frac{d^2}{2}\right) \qquad (S1)$$

$$d = \sqrt{\sum_{l=1}^{3} \frac{(\theta_l - \theta_l')^2}{\ell_l^2}} \qquad (S2)$$

where $\sigma_f^2$ is the signal variance and $\ell_l$'s are dimension-wise length scales, allowing the model to learn distinct sensitivities to each spinodoid design variable.

### S1.2 Heteroscedastic Noise Model and Training Procedure

The input-dependent noise variance $\sigma^2(\Theta)$ is modeled with the polynomial-exponential form of [1] restated here for completeness,

$$\sigma^2(\Theta) = \left[k_{al} \exp\left(\sum_{r=1}^{d}\sum_{j=1}^{p} a_{rj}\theta_j^r\right)\right]^2, \qquad p = 3, \tag{S3}$$

where $k_{al} > 0$ is a scaling factor and $a_{rj}$'s are polynomial coefficients relating the $r$-th power of cone angle $\theta_j$ to the log-noise level. This form was adopted, rather than a second latent GP over the noise variance, because it guarantees positivity by construction, reduces exactly to the homoscedastic model of Eq. (9) of the main text when $d = 0$, and yields directly interpretable coefficients: the sign and magnitude of $a_{rj}$ quantify how strongly widening $\theta_j$ suppresses or amplifies realization-induced scatter in a given stiffness component, as discussed in Section 4.2–4.3.

All hyperparameters — the GP kernel parameters governing $k(\Theta, \Theta')$ together with $\{k_{al}, a_{rj}\}$ — are estimated jointly by maximizing the log marginal likelihood of the heteroscedastic GP,

$$\log p(\mathbf{y}|\Theta) = -\frac{1}{2}\mathbf{y}^T[\mathbf{K}(\Theta,\Theta) + \sigma^2(\Theta)\mathbf{I}]^{-1}\mathbf{y} - \frac{1}{2}\log|\mathbf{K}(\Theta,\Theta) + \sigma^2(\Theta)\mathbf{I}| - \frac{N}{2}\log 2\pi\,, \tag{S4}$$

following the same one-shot joint-optimization strategy used in Ref.[1], since the training set here consists of sparse one-realization-per-point observations with no repeated $\Theta_i$ from which an empirical noise label could be constructed. The polynomial degree $d$ is not fixed *a priori* but selected, independently for each stiffness component, by comparing $ELPD_{approx}$[1], rather than $ELPD_{LOO}$, across candidate degrees $d \in \{0,1,2,\dots\}$ as defined in Section 3; this selection favors $d = 1$ for the majority of components in both the one-dimensional and three-dimensional design spaces (Fig. S1). For a small subset of components (e.g. $C_{1133}$ and $C_{2323}$), $ELPD_{approx}$ continues to increase marginally beyond $d = 1$; however, this gain is modest relative to the added risk of overfitting higher-oder coefficients to the sparse, one-realization-per-point training data. We therefore adopt $d = 1$ uniformly across all nine components following a parsimony principle, indicating that a single linear-in-exponent term is sufficient to capture the observed noise structure without overfitting the sparse training data.

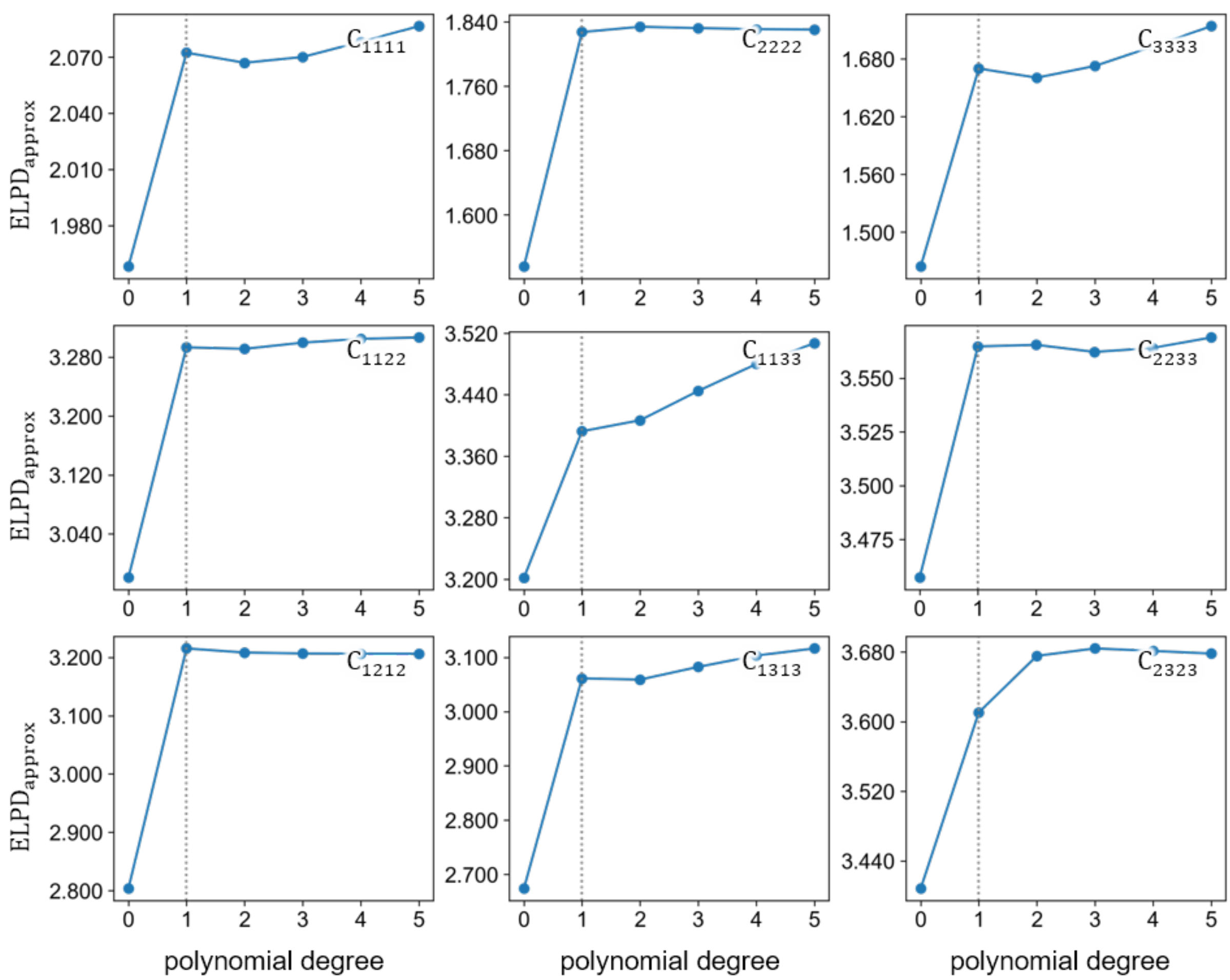

Fig. S1. $ELPD_{approx}$ as a function of polynomial degree $d$ for all nine stiffness tensor components. Most components show a clear elbow at $d = 1$ (dashed line); for $C_{1133}$ and $C_{2323}$, $ELPD_{approx}$ continues to increase marginally beyond $d = 1$, but the improvement is modest relative to the risk of overfitting given the sparse training data. $d = 1$ is therefore adopted uniformly across all nine components (Tables 1–2 of the main text), following a parsimony principle.

### S1.3 Training and Test Dataset Generation

Training inputs $\Theta_i = (\theta_1, \theta_2, \theta_3)_i$ were generated by Latin Hypercube Sampling (LHS) over the unit cube $[0,1]^3$ and mapped onto the cone-angle domain $[0°, 90°]^3$ through a power-law warping,

$$\theta_j = 90° \cdot u_j^{1.6}, \qquad j = 1,2,3, \qquad (S5)$$

rather than the uniform linear mapping $\theta_j = 90° \cdot u_j$. Because both the mean-stiffness sensitivity and the aleatoric noise magnitude are markedly larger at small cone angles than at large ones, a uniform LHS design would undersample the region of the design space that is simultaneously hardest to fit and most informative for calibrating the heteroscedastic noise model. The exponent 1.6 concentrates sampling density toward small $\theta_j$ while retaining space-filling coverage of the full domain, yielding a pool of 200 training design points.

An independent test dataset of 1,000 points was generated using standard (unwarped) LHS, uniformly distributed over the same domain $[0°, 90°]^3$ (Fig. S2), so that predictive accuracy and uncertainty calibration are assessed evenhandedly across the entire cone-angle space rather than preferentially in the densely sampled small-angle region. The sensitivity of surrogate performance to training data size was evaluated by training the heteroscedastic GPR on nested subsets of the 200-point training pool and evaluating performance on the fixed 1,000-point test data (Fig. S3). Performance plateaued beyond $N_{train} = 150$, and this training data size was therefore adopted for all analyses reported in Section 4.

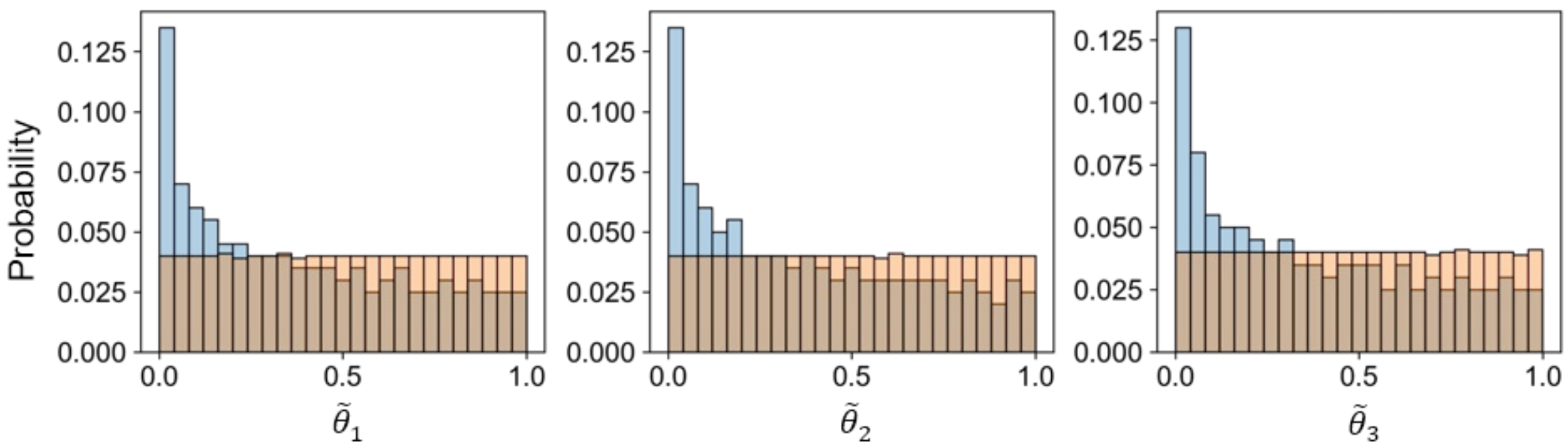

Fig. S2. Distribution of the normalized cone angles $\tilde{\theta}_j = \theta_j/90°$ $(j = 1,2,3)$ for the training data (blue, exponent-warped LHS, Eq. (S5)) and the test data (orange, uniform LHS), showing the intended bias toward small cone angles in the training data.

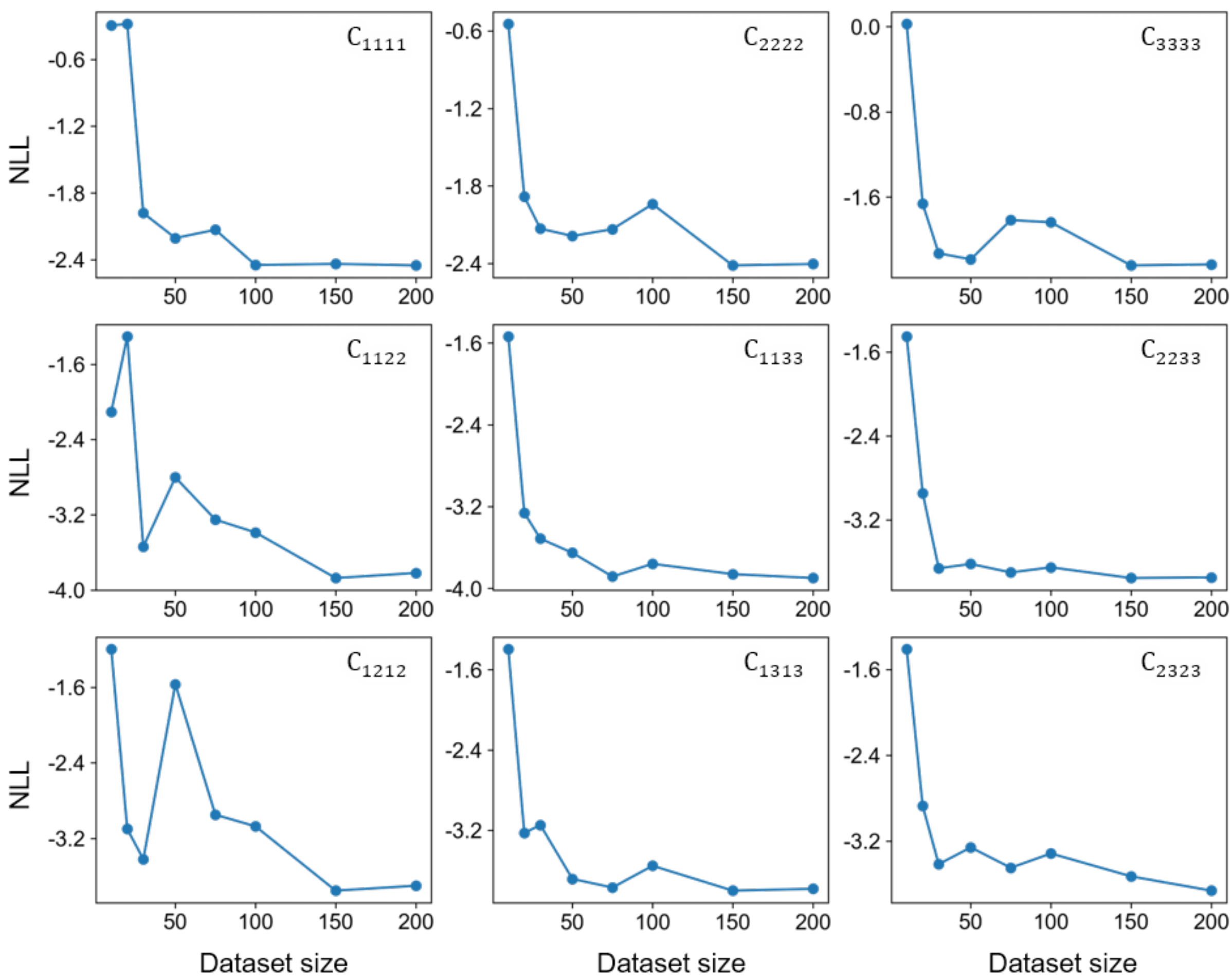

Fig. S3. Negative log-likelihood (NLL) as a function of training dataset size for all nine stiffness components. NLL plateaus beyond $N_{train} = 150$, motivating the training set size adopted in Section 4.

## S2. Additional Results

To complement Fig. 5 of the main text, which displays the predicted mean and uncertainty maps at $\theta_3 = 15°$, Fig. S4–S7 present the same $\mu(\theta_1, \theta_2)$ and $\log \sigma(\theta_1, \theta_2)$ maps for $C_{1111}$, $C_{2222}$, and $C_{1122}$ evaluated at four additional slices, $\theta_3 = 30°$, $45°$, $60°$ and $75°$, respectively. This confirms that the noise-model interpretation established for a single representative slice in the main text extrapolates consistently across the full range of the third cone angle.

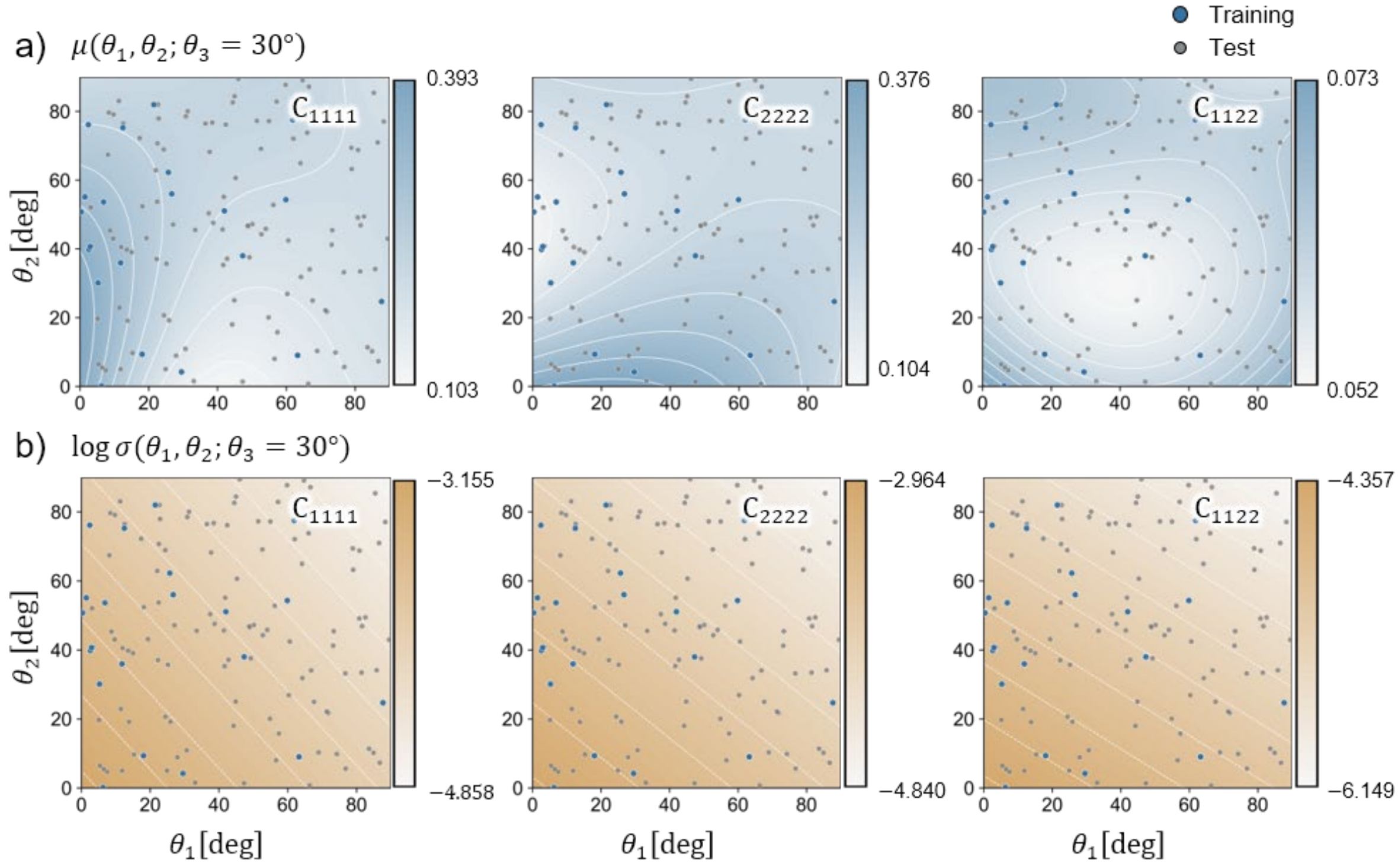

Fig. S4. Heteroscedastic GPR-predicted (a) mean $\mu(\theta_1, \theta_2)$ and (b) log-uncertainty $\log \sigma(\theta_1, \theta_2)$ maps for $C_{1111}$, $C_{2222}$, and $C_{1122}$ at $\theta_3 = 30°$, extending Fig. 5 of the main text.

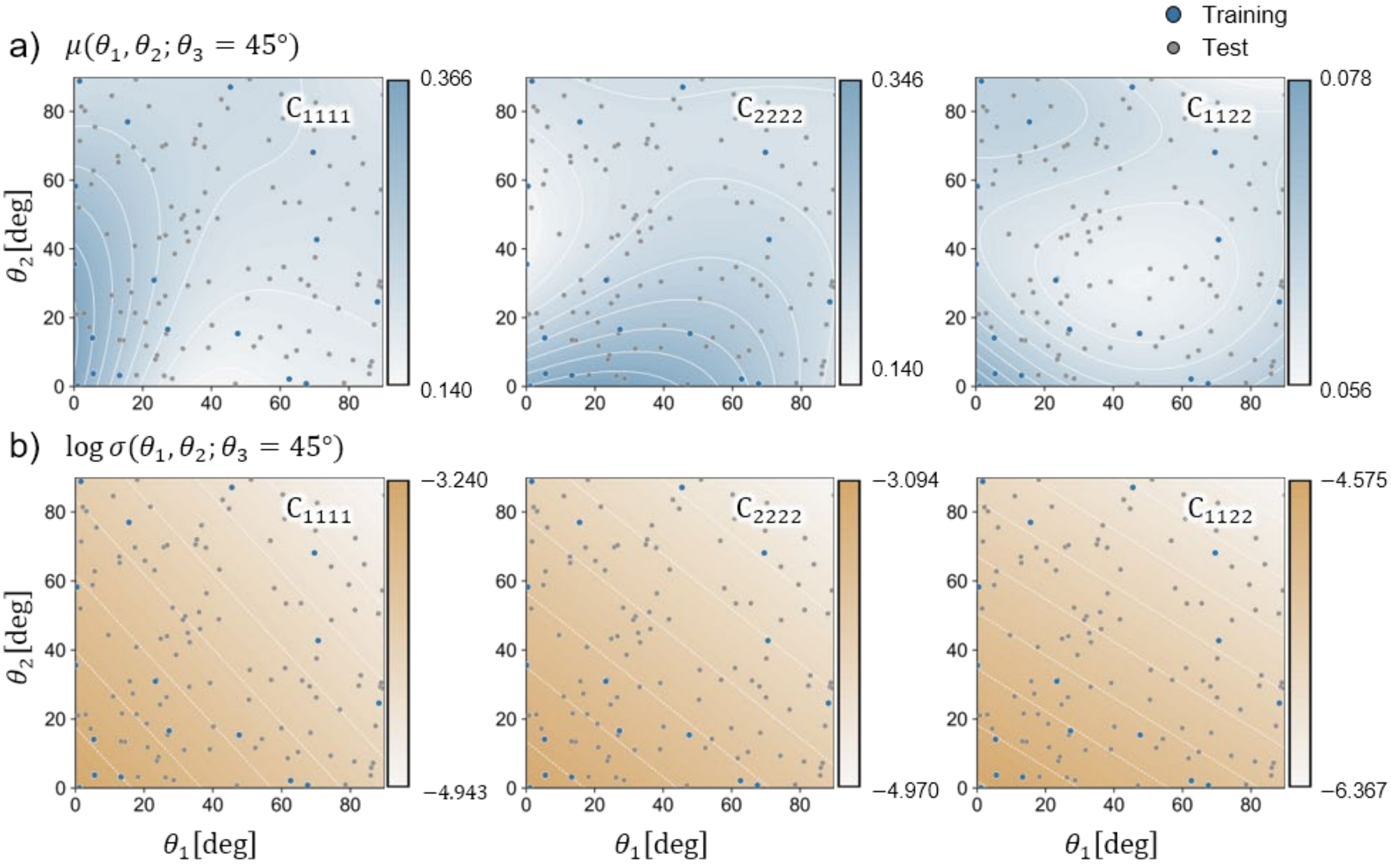

Fig. S5. Same as Fig. S4, evaluated at $\theta_3 = 45°$.

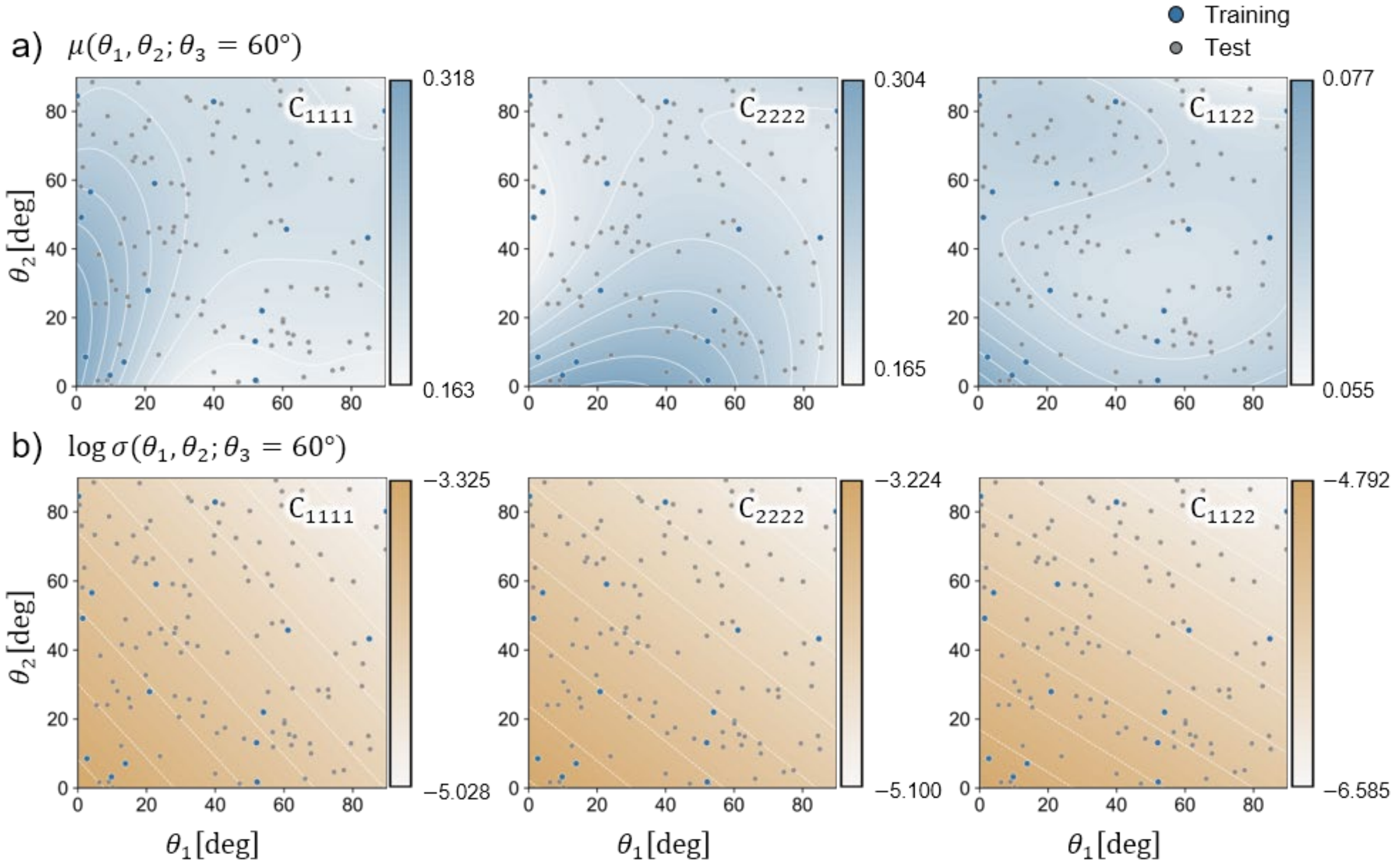

Fig. S6. Same as Fig. S4, evaluated at $\theta_3 = 60°$.

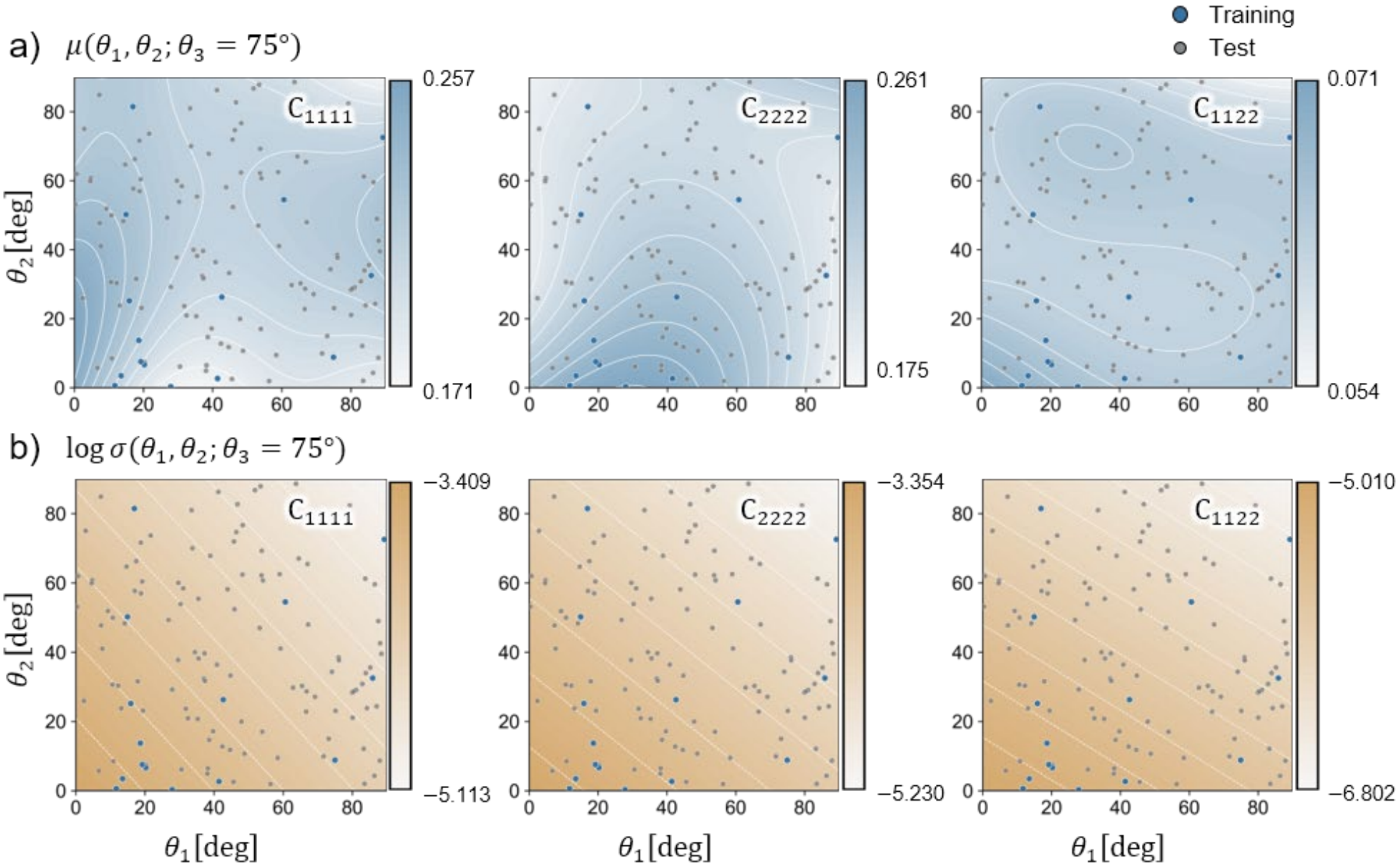

Fig. S7. Same as Fig. S4, evaluated at $\theta_3 = 75°$.